\documentclass[11pt]{article}
\usepackage{graphicx}
\usepackage{amssymb}
\usepackage{amscd}
\usepackage{mathrsfs}
\usepackage{longtable,lscape}
\usepackage{amsthm}
\usepackage{amsfonts}
\usepackage{amsmath}
\usepackage{bbm}
\usepackage{float}
\usepackage{url}
\usepackage{hyperref}
\usepackage[margin=0.5cm,font=footnotesize]{caption}
\usepackage{lineno}
\usepackage[round]{natbib}
\usepackage{appendix}
\usepackage{subfig}

\begin{document}
\title{\bf From Quantum Cognition to Conceptuality Interpretation I: Tracing the Brussels Group's Intellectual Journey}
\author{Diederik Aerts$^*$, 
Massimiliano Sassoli de Bianchi\footnote{Center Leo Apostel for Interdisciplinary Studies, 
        Vrije Universiteit Brussel (VUB), Pleinlaan 2,
         1050 Brussels, Belgium; email addresses: diraerts@vub.be,autoricerca@gmail.com} 
        $\,$ and $\,$  Sandro Sozzo\footnote{Department of Humanities and Cultural Heritage (DIUM) and Centre CQSCS, University of Udine, Vicolo Florio 2/b, 33100 Udine, Italy; email address: sandro.sozzo@uniud.it}              }

\date{}

\maketitle
\begin{abstract}
\noindent 
The \emph{conceptuality interpretation} of quantum mechanics
proposes that quantum entities have a conceptual nature, interacting with the material world through processes that are the physical counterpart of the meaning-based processes which typically occur in human cognition. This interpretation emerged from the early developments in \emph{quantum cognition}, a field that uses quantum mathematics to model human cognitive activity. It benefited from the specific approach taken by the \emph{Brussels research group}, modeling concepts themselves as quantum entities and minds as measuring apparatuses. The article sketches the essential steps of the intellectual journey going from the first applications of quantum notions and formalisms to human cognition to the proposal of a potentially groundbreaking interpretation of quantum mechanics, offering profound explanations for major quantum phenomena. This was done by drawing numerous parallels with the human conceptual domain and suggesting the existence of a level of cognitive activity that would underlie our physical reality. This means that an increased cross-fertilization between the conceptuality interpretation and quantum cognition is to be expected in the future, both of which are synergistic in furthering our understanding of the nature of reality. This is the first part of a two-part article. In the second part, which can be read independently of the first, the successes of the interpretation will be described in a more systematic way, providing a brief overview of what has been achieved so far.  
\end{abstract}
\medskip
{\bf Keywords}: 
quantum cognition, quantum mechanics, conceptuality interpretation, foundations of physics

\section{Introduction\label{intro}}

The field of research today referred to as \emph{quantum cognition} has come as a surprise. Indeed, the possibility of modeling human cognitive activity using not only the quantum mathematics, but also many of the notions belonging to the quantum formalism, followed by the discovery of the effectiveness of such modeling, was not something that could be expected. Of course, with hindsight, the success of quantum cognition appears as more natural, since both microscopic physics and the cognitive domain share, among other things, a salient feature, that of \emph{contextuality} (a term which in turn, depending on the contexts, is understood with different meanings).

The achievements of quantum cognition have led our Brussels group, over the years, to reflect not only on the technical aspects of the modeling, but also on what its success might tell us about our physical reality. Indeed, this parallel, between conceptual entities and cognitive processes on the one hand, and microscopic entities and measurement processes on the other, if it is not accidental, could tell us something fundamental about the nature of physical entities. This is perhaps closer than we might initially have inferred to that of the entities that populate our cognitive activity, quantum effects being all but absent from the macroscopic entities which we have historically interacted with. This consideration of what the success of quantum cognition might mean for the nature of our physical reality has, over time, led to the cautious formulation of an innovative interpretation of quantum physics, and possibly of relativity, which is currently being investigated in parallel with the development of quantum cognition by our group. 

Approaching this new interpretation, called the \emph{conceptuality interpretation}, without trivializing it, requires understanding the intellectual trajectory that led the \emph{Brussels research group} to its formulation, under the lead of its founder Diederik Aerts. The initial point of this trajectory corresponds to the achievements obtained in research on the axiomatic foundations of physical theories, especially quantum mechanics. Indeed, the Brussels group is the repository of what was once called the \emph{Geneva school of quantum mechanics}, which for some time was also referred to as the \emph{Geneva-Brussels school}, especially because of the fruitful collaboration between Constantin Piron and Diederik Aerts, with the latter bringing the Geneva approach not only to maturity (especially with regard to the axiomatic reconstruction of Hilbert space; see for example \citet{aertssassolisozzoproceedings2024} and the references therein) but also to reverberate in many other domains of inquiry. 

It is the trajectory of these ideas that we will try to recount in this article aware that, to better understand and further advance them, it is important to also contemplate their genesis. This is the first part of a work written in two complementary but self-consistent parts. In this first part we will concentrate on the evolution of ideas which led the Brussels group to the first pioneering studies on quantum cognition and then allowed the conceptuality interpretation to be proposed. In the second part, we will instead address in a more systematic way the quantum mysteries that the interpretation allows to elucidate \citep{aertssassolisozzo2024}.

There are numerous researchers who contributed to the initiation and development of the quantum cognition research field, in addition to the members of our group.\footnote{For practical reasons, we will not indicate which people in the Brussels group contributed to which results in the development of the quantum cognition program, considering the effort of the group as a whole in arriving at the various results. Until the early 1990s, the activities focused on the foundations of quantum mechanics. The interest in interdisciplinary research gained momentum in the 1990s, in parallel with the establishment of the interdisciplinary Center Leo Apostel, under the direction of Diederik Aerts. Many of the group's members were, and continue to be, Ph.D. students of Diederik Aerts, often transitioning to postdoctoral positions after completing their doctorates. External researchers have also joined the group regularly over the years, contributing to its ongoing interdisciplinary efforts. We mention here, in alphabetical order, the names of those who are or were members of the group, even if for some of them their activity was primarily directed towards quantum foundations, given the close connection, especially in the early years, of quantum cognition research and quantum foundations research: {\it D. Aerts, J. Aerts Argu\"{e}lles, S. Aerts, H. Amira, J. Beltran, L. Beltran, J. Broekaert, S. Bundervoet,  B. Coecke, M. Czachor, C. de Ronde, D. Deses, E. D'Hondt, B. D'Hooghe, I. Distrito, T. Durt,  L. Gabora, C. Gershenson, S. Geriente, E. Haven, F. Holik, M. Kuna, O. Leveque, C. Moreira, S. Pulmannova, J. Pycakz, M. Sassoli de Bianchi, M. Sioen, S. Smets, S. Sourbron, S. Sozzo, I. Stubbe, J. Tapia,  F. Valckenborgh, B. Van Bogaert, A. Van der Voorde, B. Van Steirthegem,  T. Veloz.}}  We mention Andrei Khrennikov and Harald Atmanspacher, for its early developments, then Jerome Busemeyer, Reinhard Blutner, Peter Bruza, Emmanuel Haven and Emmanuel Pothos, for its expansion into a full-fledged field of research, to name but a few. However, it is not the purpose of this article, which is not a review article, to relate the ideas of the Brussels group to those of other research groups. This is also because the \emph{realistic-operational approach} of our group is different from that taken by most of the authors working in quantum cognition, and it is precisely this difference that allowed the hypothesis of the conceptuality interpretation to be at some point formulated. 

This hypothesis states that \emph{quantum entities are neither particles nor waves but they are conceptual entities}, i.e., entities sharing a same conceptual nature as human concepts, interacting with ordinary material entities, like measuring apparatuses, similarly to how human concepts interact with human minds and the associated memory structures.   

The article is organized as follows. In Section~\ref{quantumcognition}, without claiming to be exhaustive, we offer our perspective on how certain guiding ideas have emerged over time, leading to the development of quantum cognition. In Section~\ref{conceptuality}, we continue our scientific narrative by exploring how the conceptuality interpretation emerged as an inevitable perspective on reality, from the inspection of the successes of quantum cognition and thanks to the specific approach taken by the Brussels group. Finally, in Section~\ref{conclusion}, we offer some concluding remarks.

\section{From quantum fluctuations to quantum cognition\label{quantumcognition}}

During the 1990s, one of us, in collaboration with one of his students, published a paper that would prove to be a pioneering work for the research area that is nowadays called \emph{quantum cognition} \citep{aertsaerts1995}. The new idea expressed in the article is that a human decision process can be described by \emph{generalized quantum probabilities}, i.e., probabilities that do violate the axioms of Kolmogorov characterizing classical probability. While the idea was innovative at the time, it was somewhat inevitable for our group, if we consider its previous insights related to the \emph{hidden-measurement approach} to quantum probability, in the 1980s. Indeed, the message that emerged from this approach is that, whenever one is in a situation where fluctuations are
present in the interaction between a measurement context and a measured entity, that is, where the context of a measurement \emph{indeterministically} affects the outcome, making it neither controllable nor predictable, one is necessarily dealing with generalized quantum probabilities \citep{aerts1993a}. And, since situations of this kind were necessarily widespread, the mechanism being very general, the expectation was that they were also abundantly present in experiments in psychology, i.e., in decision processes that were influenced by the psychological context in which they took place.

This insight was already shining through in the works of those years, for example in \citet{aertsetal1993a}, where for the first time it was clearly expressed, already in the title of the article -- Quantum structures in macroscopic reality -- that the ingredient of fluctuations in the measurement interactions could be universal in our reality. However, it was not until a 1995 article that appeared in the first issue of a new Kluwer journal, \emph{Foundations of Science}  \citep{aertsaerts1995}, which has now become a Springer-Nature journal, that it was felt that it could finally be mentioned explicitly that quantum statistics was being applied to a situation of a psychological nature. It was not obvious to do so at the time and the authors also felt that, since it was the first issue of a new journal, the reviewers would be more open than usual. This was undoubtedly a precursor article of the quantum cognition research field, which started to develop only ten years later.\footnote{Articles of this type are usually called \emph{sleeping beauties} \citep{ke2015}.}  

In those years before the international awakening of quantum cognition, we had an intent to further explore the use of quantum structures in psychology. One of us remembers well how, at that time, he would often consider examples that showed parallels with quantum measurements, like the following one. One asks a child: ``Are you hungry?'' and the child answers: ``No, I’m not hungry.'' Then one asks the same child: ``Would you like some chocolate?'' and the child says: ``Yes, I want some chocolate.'' And then one asks the same child again: ``Are you hungry?'' and the child answer is: ``Yes, I’m hungry.'' This is a good example showing how a measurement, which corresponds here to asking the question ``Do you want chocolate?'', changes the child's state in such a way that a second measurement, which corresponds here to asking the question ``Are you hungry?'', switches from the ``no'' answer to the ``yes'' answer. We are of course here considering the questions addressed to the child as the equivalent of measurements, i.e., as interrogative processes asking physical systems about their properties, with the outcomes being their answers to these questions.

In the second half of the 1990s, ideas of this kind were being cultivated in our Brussels group, qualitatively exploring possible parallels between psychological and quantum realms. Important reflections included how to make  logic operational \citep{smets2005}, as a natural extension of the group's earlier work on the axiomatic foundations of quantum mechanics, which, precisely, started from an \emph{operational} and \emph{realistic} approach  \citep{aertssassolisozzoproceedings2024}. As we shall see, such approach also proved to be fundamental to developments in quantum cognition and the conceptuality interpretation. From the perspective of these results, it appeared clear that the default state of a proposition should be one where the proposition is neither true nor false, according to what happens to a quantum property when an entity is not in an eigenstate of the \emph{yes-no measurement} operationally defining it. In other words, to paraphrase Constantin Piron, the ``if'' part is as important as the ``then'' part in quantum logic. 

It was because the group's reflections and work were also touching on issues of logic that at some point the idea emerged of putting the oldest of inconsistencies, the \emph{liar paradox}, under the quantum microscope. An operational quantum model for the  paradox, and for the thought process that it produces, was then successfully elaborated, allowing for a unitary dynamics where the mind of the person alternates between a (true) yes-state and a (false) no-state  \citep{aertsbroekaertsmets1999}. Surprisingly for us, together with the aforementioned 1995 article, also this 1999 work became one of the pioneering works of quantum cognition. 

If considerations about logic have facilitated the emergence of quantum cognition, conversely, a quantum approach to the famous liar paradox of classical logic 
may in the future open up new developments in logic, just as quantum cognition has allowed us to deepen our thinking about physical systems, as we will explain in more detail in Section~\ref{conceptuality}. A similar quantum approach could be proposed for G\"{o}del's incompleteness theorems, and this could be the beginning of a more sophisticated analysis of the so-called \emph{undecidable problems}. More precisely, the way propositions are usually classified as undecidable in mathematics should perhaps be reconsidered and, making an analogy with the difference between classical and non-classical properties in physics, it could be argued that the current focus of mathematics for propositions that can be deterministically decided would delineate only a classical domain, while most propositions would not belong to that domain, just as in quantum physics, given a state, most properties are only \emph{potential}. 

These considerations about the need to consider classical logic as just one sector within larger reasoning structures will also find expression in later work of the group, where we were able to show that the combination of concepts can be modeled quantum mechanically. More precisely, it is in the \emph{first sector} of a \emph{Fock space}\footnote{A Fock space is a special case of a Hilbert space and it is typically used in quantum mechanics and quantum field theory to represent physical situations where the number of entities is not constant.} that quantum superposition takes place, with the emergence of new meaning, whereas it is only in the \emph{second sector} of the Fock space that one can find a quantum version of classical logic \citep{aertssozzoveloz2015a}.  This suggests that there is a more primitive structure that has not been identified until now and that even Aristotle, who began formalizing human thought, overlooked it, beginning to work with the more easily identifiable second-sector structure, which then became logic as such. 

Still contained in the 1999 article on the liar paradox, there is an other aspect of the quantum model we developed that lends itself to a possible general understanding of human decisions: that of the \emph{unitary rotation} that describes the dynamics causing our human reasoning to alternatively switch between the yes (true) and no (false) answers. In a situation where one is in doubt about the available alternatives, one can observe such a rotation. In fact, when we approach a choice, the change in context that this implies may cause our choice to be all of a sudden less attractive, so other possibilities may again compete and gain preference. This mechanism is especially vivid when we find ourselves in situations of strong \emph{doubt}, causing us to constantly oscillate between the different possibilities in play. The rotation we identified in \citet{aertsbroekaertsmets1999} can be understood as an expression of this oscillation, which is usually perceptible through introspection. And it is not excluded that it is also present in very direct decisions, where apparently there is no oscillation, as in this case the rotation would just be of a very high frequency, so that it is not easily perceived through introspection. This dynamic emerged quite clearly in our model of the liar paradox and this is just one example of new ideas that can emerge when new models become available, which of course require subsequent testing. 

Let us now come to the next important episode in the development of quantum cognition research in our group. Taking advantage of its deep interdisciplinarity, and in particular the interaction with researchers having a background in psychology, the group's research began to focus on the topic of concept modeling, and more specifically on a well-known problem in that field of cognitive science called \emph{concept theory}; a problem that goes by the name of \emph{pet-fish problem}, or \emph{guppy effect}. More precisely, it has been observed by psychologists conducting research on concepts that the \emph{conjunction} of two concepts behaves peculiarly. For example, \emph{guppy} is a very typical exemplar of the conceptual combination \emph{pet-fish}, while it is not at all a typical exemplar of the concept \emph{pet} alone, nor is it a typical exemplar of the concept \emph{fish} alone \citep{osherson1981}. So, echoing our ongoing reflections on how we humans can change our opinions, the following question took shape in our group: ``Couldn't this guppy effect be another example of a genuinely quantum effect?''

In asking the latter question, one is immediately confronted with no small difficulty: one that would involve a fundamental change in vision, which today distinguishes the Brussels group from other groups working on quantum cognition \citep{aertssassolisozzo2016}. More precisely, most researchers working on quantum cognition make a connection with the quantum structures by describing the state of the human mind in a way that is mathematically equivalent to describing the state of a quantum entity. At an early stage, also in our group we began to think about the pet-fish problem in this way. Slowly, however, it became clear that a different approach was needed, and the idea of \emph{considering a concept in itself as a quantum entity} made its appearance. It turned out that this was a step of paradigmatic value, with far-reaching consequences. The conceptuality interpretation, and many new perspectives, could not have come to light without such step, which in our opinion allows for a much more fundamental approach.\footnote{More precisely, two main points stand out in our approach: (i) the states of conceptual entities express their \emph{modes of being}; (ii) in a cognitive experiment, participants act as a measurement context, changing these states. Consequently, a state, represented in a quantum approach by a unit vector in Hilbert space, does not reflect subjective beliefs, which are rather embedded in the interaction between the cognitive situation and the participants. Our operational quantum approach thus maintains a realistic stance, which distinguishes it from approaches that view the quantum state purely as a belief. Of course, interpreting the quantum state as a `state of belief' or as a `state of a conceptual entity' is to some extent just a matter of philosophy, so our group and the other quantum cognition groups are all working on the same research program. But interpretation can also shape methodology, as different views influence how approaches are developed. If we see particular value in our operational-realistic approach, which balances idealist and realistic interpretations, it is because it has allowed us, for example, to use quantum superpositions to model the states of conceptual combinations. This way of representing combined concepts as entities in linear superpositions of individual states, which is also how emergence is captured in the physical processes described by quantum mechanics, would in our view have been much harder to achieve if the focus had been solely on belief states \citep{aertssassolisozzo2016}.}

In the first phase of these reflections, we managed to sketch a scheme able to make a connection with the quantum axiomatics of a so-called \emph{State Property System}, a mathematical structure that arises when a physical entity is assumed to be, at any time, in a well defined state, with its properties being either \emph{actual} or \emph{potential}, worked out in the early 1980s \citep{aerts1982} and refined in the late 1990s early 2000s \citep{aertsdeses2002,aertscolebunders2002,aertspulmannova2006}. What became clear is that something needed to be added to this formalism, if used to describe concepts. It turned out that, although for a quantum entity the notion of \emph{context} coincides with how a \emph{property} is operationally defined, for concepts we quickly found examples where the two notions were not equivalent. Thus, we were obliged to define contexts separately from properties. In this way, we came to the definition of a \emph{State Context Property system}, abbreviated to \emph{SCoP}, which we used to describe the contextual manner in which concepts are evoked, used, and combined to generate meaning \citep{gaboraaerts2002}.

Another important step in the Brussels research group, as part of its analysis of conceptual entities, was to begin using the World-Wide Web as an investigative tool, since it contained, within its huge corpora of documents, countless traces of human cognitive activity, thus allowing to highlight how different words, and combinations of words, expression of different concepts in different states, could bound together through the invisible substance of \emph{meaning}. In this way, it was later possible to also retrieve the guppy effect underlying the pet-fish problem directly on the World-Wide Web \citep{acds2010}. Here the idea was that, if one uses a search engine to enter terms that correspond to typical questions one would ask in psychological experiments, then, as a consequence of the fact that the Web contains texts that are written by people, and therefore represent their minds, the results of those searches are expected to contain webpages correlated with the answers that test subjects would typically give in such psychological experiments.  

Let us provide an example. Suppose one collects the webpages that contain the word \emph{Bird}.\footnote{We use concepts and the words that name them interchangeably throughout this article.} Then, those webpages will also more often contain, statistically speaking, the word \emph{Wing}, compared to generic pages across the Web. This simply because if \emph{Bird} is contained in a webpage, there is a greater than random chance that the text on that page is about birds, and if that is true, there is a greater than random chance that it contains the word \emph{Wing}. In other words, words associated in meaning are more likely to occur together on the same webpage. In analyzing these possibilities, and again thanks to the interactions happening in a research group formed by people with multiple expertises, it did not take long to realize that the idea of identifying \emph{meaning connections} on the Web was actually not new. In fact, as early as the 1990s, computer scientists developed theories about meaning connections in a research area called \emph{semantic analysis}, and an approach that attracted all our attention was called \emph{latent semantic analysis}, or LSA in brief. 

As we delved deeper into the topic, our amazement grew more and more, as work of computer scientists on semantic analysis and LSA revealed the existence of evident links to quantum mechanics. And it was also clear that the researchers working in those areas were unaware of these structural similarities. A short but incisive article was written to emphasize the structural connections between major approaches to semantic analysis and the Hilbert-space formalism \citep{aertsczachor2004}, which immediately caught the attention of some researchers. Of particular interest was our observation that computer scientists did not use in their studies the \emph{tensor product} of vector spaces, which instead plays a fundamental role in quantum mechanics. 

When our article had just been published, we were contacted by Dominic Widdows, a researcher at Stanford University, who appreciated our comment on the failed use of the tensor product in LSA and asked us if we had any concrete results specifically related to it. At the same time it became clear that he himself had been working from a similar inspiration to ours, and more specifically had replaced the negation of classical (Boolean) logic with the orthogonality relation of quantum logic in information retrieval systems \citep{Widdows2003}. A few months after Widdows had contacted us, Keith van Rijsbergen's book was published, also linking quantum mechanics and information retrieval \citep{Rijsbergen2004}. These three events in the same year 2004, clearly independent from each other according to the chronology of the publications, initiated a new field of research, similar to quantum cognition, but with a focus on how human cognition is structurally present in texts on the World-Wide Web. It has been called \emph{quantum information science},\footnote{Quantum information science must be distinguished from the \emph{theory of quantum information}, which does not aim at identifying quantum structures within classical information but, rather, at exploiting the quantum information stored in micro-physical entities to perform new types of computational tasks.} with subfields focused on \emph{information retrieval} and \emph{natural language processing}.

Let us briefly explain how the LSA technique is used in practice by computer scientists. Starting from a corpus of documents, a matrix is typically formed out of it, with the words corresponding to the rows and the paragraphs (or other portions of the considered documents) to the columns. Then, the matrix is filled with the numbers of times the different words occur in the different paragraphs. Since one is dealing with a corpus of many documents that contain many different words, the obtained matrix is generally very large and most of its elements are zero. Indeed, a random paragraph will generally not contain a random word. In the 1990s, computer power was not yet sufficient to handle such large matrices, so a technique was used to handle them more efficiently, called \emph{singular value decomposition}, a generalization of the typical decomposition of a square matrix into a diagonal matrix. More precisely, one writes the matrix in question as the product of three matrices, the first and last being orthogonal matrices and the middle one a square matrix. This middle matrix is then diagonalized, and the small eigenvalues are omitted, so the rank of the matrix is lowered. If the original matrix is then recalculated, one obtains a new matrix where most of the previous zeros are replaced by non-zero values. This means that words that do not occur in certain paragraphs are now assigned  non-zero weights for those paragraphs, which explains the term ``latent'' in the LSA designation.

When dealing with \emph{word-document matrices}, it was natural for people in LSA to use the techniques of linear algebra, hence to work with vectors. Of course, if one considers very large corpora of documents, one quickly comes to consider very large matrices, which will not be square matrices as there are much more documents (or portions of documents) than words, if one considers a large collection of documents. When trying to diagonalize pieces of such non-square matrices, the diagonalization created eigenvector directions in a real vector space, which were interpretable as \emph{meaning directions}. When doing so, similarly to how one would eliminate noise in quantum mechanics, the rank of the matrices were lowered by dropping the small eigenvalues, keeping only the dominant meaning directions. And when going back to the initial matrices, one would discover connections between documents and terms, through the linear vector space, even when a word would not explicitly appear in a given document. 

More specifically, the similarity between two vectors, measured by calculating their scalar product (the \emph{cosine} of their angle), would then provide a measure of the \emph{closeness in meaning} of a certain word with a certain document, even when that term did not appear in the document in question, i.e., when it only appeared in a \emph{latent} way. This description was very similar to that of the collapse of the wave function of quantum physics, replacing the notion of \emph{latency} by that of \emph{potentiality}. But apart from terminological issues, there was an important structural difference with respect to the quantum formalism: they were only working in real vector spaces, not in complex vector spaces. Also, they didn’t know about the \emph{tensor product}. Probably the reason for this is that a tensor product will not appear naturally as a structure to a computer scientist using linear algebra and vector spaces. The notion of tensor product was first introduced in quantum mechanics by von Neumann in 1932 \citep{neumann1932}, in his Hilbertian formalization of the theory, but physicists were kind of forced to introduce it to be able to represent composite quantum entities and the associated phenomena of quantum entanglement, coupling of spins, etc.

A tensor product, however, is also fundamental to represent how concepts combine, for instance using the structure of a Fock space, as we have already mentioned. Let us explain this with some more details. Suppose one combines the concept \emph{Fruit} with the concept \emph{Vegetable}, for example considering the conceptual disjunction \emph{Fruit or Vegetable}. One can then look at single exemplars for such combination, like \emph{Mushroom}, and check what the weights of these exemplars are with respect to \emph{Fruit or Vegetable} (see also Section~\ref{conceptuality}). When modeling these weights, one needs to use the \emph{superposition principle}, i.e., the linearity of the state space. In other words, one can remain in this case in the first sector of Fock space, without the need of introducing the tensor product structure. The situation is however different when taking a conceptual combination like \emph{The Animal Acts}, and then consider exemplars that are also combinations of concepts, like \emph{The Horse Whinnies}, \emph{The Bear Growls}, \emph{The Bear Whinnies}, etc. The tensor product then immediately appears as the natural structure to consider, as is the case in physics when dealing with entities that are combinations of sub-entities. Hence, we are now in the second sector of the Fock space. The general structure one needs to consider is therefore the Fock space structure formed by the direct sum of these two sectors. The first one is where one looks at exemplars of a single conceptual entity, like in the \emph{Fruit or Vegetable} example, and a second one is where one looks at exemplars that consist of two conceptual entities, like in the  \emph{The Animal Acts} example \citep{aertssozzoveloz2015a,aertssozzoveloz2015b}.

So, computer scientists, when looking at matrices and diagonalization techniques remained in the first sector of the Fock space and, without really being aware of that, they were actuallygġ looking into superpositions of a quantum type. And since it is in the first sector of Fock space that one can find the strongest quantum effects, even stronger than entanglement, this explains why the LSA people could still construct a strong formalism. Of course, another crucial step that was not taken by computer scientists is the passage to complex numbers. By working in real vector spaces, the full power of the quantum machinery was missing, even in the first sector, which is why in our 2004 article we tried to explain how to bring the full quantum structure into the game, i.e., how to reformulate semantic analysis as a Hilbert-space problem \citep{aertsczachor2004}. These were the years when both \emph{quantum cognition} and \emph{quantum information science} began to become well-defined fields of research, thanks also to the efforts of people like Peter Bruza, who organised regular meetings under the heading `quantum interaction', with proceedings \citep{bruzaetal2007,bruzaetal2009}, and Jerome Busemeyer, who wrote the first book on quantum cognition with Bruza in 2012 \citep{BusemeyerBruza2012}. Other scientists we can mention (in addition to those in our group) who played a role in the flourishing of quantum cognition are Andrei Khrennikov and Harald Atmanspacher. 

We believe that quantum cognition touches much deeper aspects of reality than is commonly imagined, and the recent results on the connection between the linguistic \emph{Zipf's law} and the \emph{Bose-Einstein statistics} (more on that in Section~\ref{conceptuality}) confirms this in our opinion \citep{aertsbeltran2020,aertsbeltran2022}. What this deeper level would be telling us is that there is much more cognitive activity in the physical world than we initially expected, and it is in following this insight that an interpretation of quantum mechanics took shape over time, with a first publication in 2009, called the  \emph{conceptuality interpretation} \citep{aerts2009,Aerts2010a,Aerts2010b,Aerts2013,aerts2014,aertsetal2020,aertssassoli2024b}. It is not impossible that it will be through it that quantum cognition will be explored more extensively in the future, provided that people will be ready to accept the radical paradigm shift it entails.

\section{From quantum cognition to cognitons\label{conceptuality}}

The intellectual gestation of the conceptuality interpretation was carried out initially by one of us (Diederik Aerts), who then over time engaged several members of the Brussels group, who fascinated by its perspective contributed to its development. The interpretation emerged in 2005, during a period of deep reflection following the mentioned pioneering articles about quantum cognition and quantum information science \citep{aertsaerts1995,aertsbroekaertsmets1999,gaboraaerts2002,aertsczachor2004,aertsgabora2005,aertsgabora2005b}, but its first publication was only in 2009 \citep{aerts2009}. Now, before being able to fully formalize it, several modeling possibilities were on the table, e.g., whether one should model concepts as subspaces or as vectors of a Hilbert space. It is also worth observing that applying the quantum formalism the right way, to explain the data of experimental psychology, was not at all a straightforward operation in the beginning, particularly so if one was aiming at identifying non-ad hoc and possibly universal approaches.

Being able to understand how the conceptuality interpretation emerged from these ongoing reflections on quantum cognition is very important from our perspective. Indeed, while it is easy for people in our group who worked on it to recognize that it truly holds a powerful explanation for the strange behavior of quantum entities, and might well turn out to be the only interpretation able to do so, it is also very easy to misunderstand it and grasp it only superficially, also because when one starts to explore it more deeply,  it is quite a leap paradigmatically speaking: it completely changes how one sees the entire physical reality.

A crucial role in its formulation was played by Hampton's data on the \emph{disjunction} and \emph{conjunction} of two concepts, such as the example we have 
mentioned in Section \ref{quantumcognition} regarding the combination  \emph{Fruit or Vegetable} \citep{hampton1988a,hampton1988b}, as it allowed to approach the mystery of the \emph{double-slit experiment} from a completely new perspective. Hampton's probabilistic intuition made him correctly identify that what he was measuring was a violation of the axioms of Kolmogorov of classical probability. However, he did not consider that such violation could indicate the presence of quantum probabilities. In other words, he was not analyzing his data with a knowledge of the difference between classical and quantum probability, which on the other hand was part of the knowledge base of our group. In these experiments, the \emph{membership weight} of an exemplar with respect to different concepts and their combinations was measured, by having subjects providing estimates on a so-called \emph{Likert scale}.\footnote{In addition to measuring the membership weight on a Likert scale, Hampton also directly measured the probability of a subject judging a particular exemplar as a member of a given concept. In fact, it was this direct probabilistic measurement that more convincingly indicated the presence of non-classical, and possibly quantum, probability.}
Such estimates are of course very complex operations to perform from a cognitive perspective, but no doubts the estimation of the membership weight
of an exemplar is strongly correlated to the frequency with which the exemplar is chosen as member of a given concept. In other words, although indirectly, Hampton performed yes-no measurements, and the calculated probabilities were found to violate the rules of classical probability. 

For example, Hampton measured the membership weights of \emph{Mint} with respect to the concepts \emph{Food}, \emph{Plant} and their conjunction \emph{Food and Plant}, finding that the membership weight of \emph{Mint} with respect to \emph{Food and Plant} was strictly higher than the membership weight of \emph{Mint} with respect to both \emph{Food} and \emph{Plant}, an effect that Hampton called \emph{double overextension}. Since the membership weight of an exemplar with respect to a given concept is the probability that the exemplar is judged as a member of that concept, the empirical effect identified by Hampton was a significant violation of the rules of classical probability \citep{hampton1988a}. Analogously, Hampton measured the membership weights of \emph{Sunglasses} with respect to the concepts \emph{Sportswear}, \emph{Sport Equipment} and their disjunction \emph{Sportswear or Sport Equipment}, finding that the membership weight of \emph{Sunglasses} with respect to \emph{Sportswear or Sport Equipment} was strictly lower than the membership weight of \emph{Sunglasses} with respect to both \emph{Sportswear} and \emph{Sport Equipment}, again a substantial violation of the rules of classical probability, called by Hampton \emph{double underextension} \citep{hampton1988b}.

Considering these results, the idea to represent concepts by \emph{closed subspaces} of a Hilbert space, in analogy with how properties are represented in quantum logic, was not anymore tenable, as then it was easy to show that the violations identified by Hampton could not occur. Indeed, the conjunction of two concepts would then be represented by the intersection of the closed subspaces representing each of them. But the quantum collapse probability of a unit vector to this intersection is always smaller or equal to the quantum collapse probability of this vector to each of the closed subspaces, which means that the classical probability rule with respect to the conjunction would be satisfied. Similarly, the quantum collapse probability of a unit vector to the closed subspace linearly and topologically generated by both subspaces, representing the disjunction, is always greater or equal to the quantum collapse probability of this vector to each of the subspaces, which means that also the classical probability rule with respect to the disjunction would be satisfied. So, starting from quantum logic, the obvious thing one would be tempted to do, to capture the non-classical structure in the data measured by Hampton, cannot work, because a concept is not fully defined by its properties, as Aristotle believed, and this explains why in the end it was convenient to use vector-states to represent concepts, in accordance with what was done by computer scientists. 

Hence, following the quantum formalism, concepts were considered to be the equivalent of the microscopic quantum entities, whereas minds, also to be understood as memory structures sensitive to meaning, the equivalent of the measuring apparatuses. And it was important to also identify the existence and influence of different contexts for a concept, the \emph{indeterministic} ones, corresponding to measurements, i.e., to decision making situations where the actual breaks the symmetry of the potential, and the \emph{deterministic} ones, corresponding to the different possible expressions that can change the perceived meaning of a concept, hence its state. Just to give an example of a deterministic context, think of the meaning carried by the concept \emph{Fruit}, in its neutral \emph{ground state}, when not in combination with other concepts, and the different meaning it carries when in the \emph{excited state} specified by the combination \emph{Juicy Fruit}, which will lend itself to the selection, with higher probability, of exemplars of fruits having an increased content in water. And of course some contexts will be more able than others to bring a concept into a more \emph{concrete} state, that is, closer to a condition that we usually would refer to as an \emph{object}.

The concept \emph{Thing} can be considered to be very \emph{abstract}. Indeed, any entity in our world can be considered to be a good representative of a \emph{Thing}. But if we say \emph{That Very Tall Thing that is a Symbol of one of the Greatest European Cities}, it is already a much more localized state in the conceptual landscape, as only a few entities are good examples of it. And if we say  \emph{That Very Tall Metallic Thing Symbol of Paris}, then there is only one entity, hence the concept  \emph{Thing}, in the state \emph{That Very Tall Metallic Thing Symbol of Paris}, comes into correspondence with an object in our physical universe, the Eiffel Tower. 

These considerations lead us straight to an understanding of \emph{Heisenberg's uncertainty principle} as an ontological statement. Indeed, if it is true that quantum entities, like photons, electrons, atoms, etc., are \emph{meaning entities}, as the concepts of our human language are, then they can also be in states of varying degrees of abstractness, or concreteness. But we also know that a concept cannot be maximally abstract and maximally concrete at the same time, and this tradeoff between abstractness and concreteness would be exactly what gives rise to Heisenberg’s principle. The most abstract concepts, like \emph{Thing}, that are able to collapse onto anything, would then be the equivalent of the \emph{plane waves}, whereas the most concrete concepts, associated with the spatiotemporal objects, would be the equivalent of the \emph{delta functions}. 

Before adding more, a warning is necessary. When we say that quantum entities are like human concepts, what one must understand is that they are of a conceptual kind, i.e., that they share with human concepts a similar nature, but we should avoid undue anthropomorphizations. In other words, we can use our intuitions about human concepts as a guide for our reasonings, but the physical entities, whose behavior we want to explain, are not human concepts: they would be entities belonging to a very different ``cultural layer'' of our reality. 

That being said, even more striking is how the conceptuality hypothesis allows one to understand the notion of \emph{indistinguishability}. Note that there is an important distinction between \emph{identicality} and \emph{indistinguishability}. Even if two objects are identical, such as two marbles having the same color, size, etc., since they are spatial objects, we can always put them in different places and, by doing so, we can distinguish them. Thus, being identical is not the same as being indistinguishable. Indistinguishability requires identicality, but not the other way around. Now, if we say \emph{Seven Cats}, then, as concepts, they are not only identical entities, but also genuinely indistinguishable entities, which is not the case if we considered them as objects. 

To better explain this different behavior between objects and concepts, suppose one goes to a farm, with seven baskets, and ask the farmer to randomly fill them with cats and dogs from the farm. The probability that seven cats (or seven dogs) are selected is then much smaller than that of having four cats and three dogs in the baskets. This probability difference is computable by using the \emph{Maxwell-Boltzmann} statistical distribution. For example, if the farmer's probability of selecting a cat is equal to that of selecting a dog, there is $35$ times more chances of having four cats and three dogs in the baskets than seven cats. Indeed, $3!=6$,  $4!=24$, $24\cdot 6=144$, $7! = 5040$, and dividing the latter by $144$ we obtain $35$, which is the \emph{multiplicity} of a state of three identical dogs and four identical cats, expressing all the different ways one can put them in seven distinct baskets. Instead, for seven identical cats (or seven identical dogs) the multiplicity is just $1$, as there is only one way to put them in the baskets.

On the other hand, if someone is asked not to select `seven physical cats' or `four physical cats and three physical dogs', but to choose between  \emph{Seven Cats} and \emph{Four Cats and Three Dogs}, in a purely conceptual way, that is, not as actual cats and dogs of the farm, to be put into actual baskets, but as two different examples of  \emph{Seven Animals}, expressed at the linguistic level only, then the states describing their meaning will have the same multiplicity of $1$. So, the purely conceptual situation is not of a Maxwell-Boltzmann type, where entities are identical but still distinguishable. It is more like a Bose-Einstein type of situation, where entities are genuinely indistinguishable. If this is true, it has to be testable, and this was indeed initially confirmed back in 2009, when using a search engine on the World-Wide Web, to collect data \citep{aerts2009,aertsetal2020}.

In more recent times, the hypothesis that indistinguishability would be a consequence of the conceptual nature of quantum entities was studied much more carefully, by showing that one can analyze a text of human language telling a story (formed by the combination of numerous words, hence concepts) by introducing \emph{energy levels}, with the word appearing most frequently in the text occupying the ground state, then the word appearing second most frequently occupying the first energy level, and so on. By doing so, one finds the remarkable result that it is the \emph{Bose-Einstein statistics} that can efficiently model the distribution of words in the story and, even more remarkable, that Zipf's law in human language is nothing but an expression of such quantum statistics \citep{aertsbeltran2020, aertsbeltran2022}. Now, in the same way a text can be interpreted as a Bose gas in a state close to a \emph{Bose-Einstein condensate}, reversing the perspective one can understand the behavior of a gas of bosons near absolute zero temperature by viewing it as a collection of conceptual entities that connect through meaning to form what in our human language we would call a \emph{story}. This is the reason we introduced the notion of \emph{cogniton}, as the \emph{quantum of human thought}, which plays the same role, within our human language, of a bosonic entity in a quantum gas. And of course, this
quantum theoretical statistical analysis of texts, certainly interesting from the perspective of modeling human language per se, brings strong new evidence for our conceptuality interpretation of quantum mechanics. 

From a historical point of view, the next explanatory success of the conceptuality interpretation was the ability to explain the interference effects described by the superposition principle, as manifested, for example, in typical double-slit experiments. Considering the cognitive equivalent of these experiments which were carried out by Hampton \citep{hampton1988a,hampton1988b}, the difficulty faced here was that of being able to understand what was going on in the subjects' minds when overextension or underextension effects appeared, and how this could be translated to explain a double-slit experiment in the physics lab. Considering Hampton's example about the disjunction \emph{Fruit or Vegetable}, a fundamental insight was to realize that  it could be viewed in two ways. The first was to really consider it as the classical logical disjunction of two concepts, hence, the concept \emph{Fruit} \emph{or} the concept \emph{Vegetable}. 
The second was to view it as a brand-new concept, created by their combination, for which genuine new properties could emerge. Focusing on this second way, it became evident why an exemplar like \emph{Olive} would have an overextended probability to be classified as a member of \emph{Fruit or Vegetable}. Indeed,  \emph{Olive} is not considered to be a member of \emph{Fruit}, nor a member of \emph{Vegetable}, but precisely because of that, it is more easily described by the combination \emph{Fruit or Vegetable}, as being better able to express the \emph{doubt} about its classification. On the other hand, an exemplar like \emph{Elderberry}, which was considered a member of \emph{Fruit} but not a member of \emph{Vegetable}, it was not a good example of a situation of doubt as to whether it belongs to one of these two categories, so it was not well described by the  \emph{Fruit or Vegetable} emergent meaning, whence the underextension effect.

When this kind of understanding was transferred to a double-slit experiment as typically performed in a physic's lab, it allowed us to interpret it in a whole new way, imagining that the detector screen was the equivalent of a cognitive entity using impacts to ``write down'' answers to specific questions. When both slits are open, the question, if we translate it in our human language, conveys the following meaning: ``What is the best example of an impact that can express a situation of doubt about the slit from which the particle emerged?''. It is then not difficult to realize that it is exactly halfway between the two slits that the most likely answers will be found, as these are the locations best able to express that situation of doubt. And that would be the reason why it is precisely there that one finds the fringe of greatest intensity, in a region where one would classically expect to find only very rare traces of impact. 

Thus, as it was the case with the difference between `physical cats and dogs' and `conceptual cats and dogs', reasoning on the level of meaning expressed by the question addressed to the ``screen cognitive entity,'' then trying to give answers that best express that meaning, is not the same thing as reasoning in terms of corpuscular material entities actually moving in space. Because it is in only when we move to the more abstract (non-spatial) conceptual layer that we can reason on the level of possibilities, rather than on that of actualities, and this is what would make all the difference between quantum and classical processes. To put it another way, the change in perspective is to not try to determine from which slit the quantum entity would emerge, but what are the locations on the screen that would be good representatives, as localized impacts, of situations of doubt about from where it emerged. According to the conceptuality hypothesis, this is exactly what a detection screen does, when it randomly but meaningfully actualizes impacts. 

An important observation, when reasoning in this way, is that the conceptuality interpretation is constantly reinforced by the successes of quantum cognition, just as the latter is reinforced by the successes of the conceptuality interpretation. Indeed, while considering that there are considerable differences between the human conceptual domain, developed only recently in our evolution, and the hypothetical matter-energy cognitive domain, which would be older and therefore more structured, the expectation is that numerous meaningful parallels are to be found between these two domains. For example, if it is possible in a physics laboratory to highlight interference patterns, the same should be possible in the psychology laboratory. To verify this, all that was needed was to analyze Hampton's data in a truly quantum way, as if it were data from an actual double-slit experiment, placing the different exemplars on a two-dimensional plane acting as a detector screen. A possibility was then to use two $2$-dimensional Gaussian wave packets to model the  \emph{Fruit} and  \emph{Vegetable} data, then consider their superposition to model the  \emph{Fruit or Vegetable} data, and the result was a surprising (birefringence-like) interference pattern \citep{aerts2009,aertsetal2020}.

The conceptuality approach to the double-slit experiment also allowed us to understand why interference patterns can form impact after impact, one quantum entity at a time. It is in fact the meaning expressed by the overall experimental setting, and the interrogative context associated with it, that produces the observed effects, by guiding the quantum entities to the different possibilities, that is, to the different possible answers, preferring of course those that best represent the conceptual structure associated with the measuring apparatus. And in this way one can also easily explain the so-called \emph{delayed-choice quantum eraser experiments} \citep{Yoon2000}. Indeed, erasing information that already existed amounts to changing the conceptual structure at the level of the measuring apparatus, hence the very nature of the questions asked, and it is therefore to be expected that these peculiar effects are observable and real, which would be difficult if not impossible to understand if one continues to think of quantum entities as small objects or waves.

\section{Conclusion\label{conclusion}}

There would be much to add to explain the various subtleties that the conceptuality interpretation brings to bear and how it has been able, over time, to explain many other quantum effects \citep{aerts2009,Aerts2010a,Aerts2010b,Aerts2013,aerts2014,aertsetal2020,aertssassoli2024b}. As we have partly explained, \emph{quantum superpositions} can be understood as descriptions of the emergent meanings arising from conceptual combinations; \emph{quantum correlations} as the manifestation of the connections of meaning existing between different concepts, able to violate Bell's inequaltities \citep{as2011,aabgssv2019};  \emph{quantum complementarity} as a tradeoff between the abstract and the concrete, relative to a given experimental context; \emph{quantum non-locality} as the manifestation of the more abstract, non-spatial states of conceptual entities \citep{sassoli2021}. Also the phenomenon of \emph{quantization of quantum observables} can be explained as an equivalent of \emph{categorical perception} \citep{aertsaerts2022}, i.e., the typical warping of conceptual perception where stimuli clump together to form quanta, leading to a discretization of a dimension. And it is also possible to provide a detailed description of what happens during the collapse of the wave function, describing it as a \emph{tension-reduction} cognitive process \citep{aertssassoli2014,aertssassoli2015}. As far as relativity is concerned, the conceptuality interpretation can also explain the phenomenon of \emph{time dilation}, when one compares different cognitive pathways starting from a common hypothesis and reaching a same conclusion \citep{aertsetal2020,aerts2018,aertssassoli2024a}. 

To explain all this is however not the scope of this first part of a two-part article, where we tried not so much to be comprehensive, as to sketch a trajectory of ideas that led to the development of the conceptuality interpretation, starting from the development of quantum cognition and considering the mutual fertilization between these two perspectives on reality. With quantum cognition we tried to better understand human cognition by looking at it as if it were a quantum phenomenon. With the conceptuality interpretation we tried to better understand quantum physics and relativity by analyzing them as if they were telling us something about hidden cognitive processes. And if the latter are real, this will have a significant impact on the way we understand reality as a whole, for example the evolution of life and complexity in a broad sense \citep{AertsEtal2011,aertssassoli2018}. In the second part of this article we will provide a more systematic, but nevertheless still concise, overview of the conceptuality interpretation, which will constitute a complementary reading to what was here more narratively exposed \citep{aertssassolisozzo2024}.

To conclude, we briefly mention an important difference that exists between the human conceptual realm and the hypothetical physical conceptual realm. In human language, a complication arises due to the fact that it contains two distinct directions to go from the abstract to the concrete. The first corresponds to the creation of concepts by abstracting them from objects. The second considers the single-word concepts as being the most abstract ones, and only when combined together, in forming stories, they become increasingly concrete. Computer scientists coincidentally stumbled upon this second direction, which is the more universal one, while psychologists and semiologists have mostly dealt with the first, more influenced by our specific situation of living and thinking beings evolving on the surface of planet Earth.

The reason why it is important to be aware of these two directions of going from abstract to concrete, let us call them the \emph{universal} and the \emph{parochial}, is that in the human conceptual realm the latter produces a distortion of the more organized structure generated by the former. The parochial line exists because we humans made contact with the entities belonging to a different conceptual realm, our physical domain, which according to the conceptuality interpretation is also conceptual in nature. On the other hand, the universal line comes from the development of our language, when for instance we create meaningful stories formed by a great number of words, and the same probably happens in the physical domain, with atoms and molecules, when they combine to create a macroscopic object, which would be the equivalent of a story. But in this case the \emph{one-word} building blocks are not the words of our human language, but those of a much more ancient protolanguage that we humans are unable to speak. 

Again, if it is important to use human cognition as an example of a cognitive domain that can guide us in exploring new possible cognitive domains in reality, like the quantum one, equally important is not to confuse the different domains, recognizing the differences that distinguish them. In addition to the one already mentioned, of the presence of a parochial line that must be added to the universal one, another important difference is that there is much more coherence (meaning) in the human cognitive domains than in the macroscopic environment in which we humans have evolved, confronted with the decoherence processes that the photonic bombardment produces, destroying most of the quantum phenomena that can be observed only under controlled conditions in a laboratory. In that sense, the cognitive environment in which our human language unfolds and evolves is much colder compared to the environment where the biological life unfolded and evolved, on the surface of this planet. Regardless of these differences, the hypothesis is that humans, with their language, have simply re-exploited that same potentiality built in reality, in order to interact with other human beings in a cognitive way. The conceptuality interpretation assumes that this potentiality was also exploited very long ago, which would explain the existence of the \emph{boson-fermion duality}, the bosons being the words (the cognitons) of the protolanguage and the fermions the building blocks of the protominds that enabled the dialogue that created the complex physical structures we are able today to contemplate and analyze.

\section*{Acknowledgements}
This work was supported by the project ``New Methodologies for Information Access and Retrieval with Applications to the Digital Humanities'', scientist in charge S. Sozzo, financed within the fund ``DIUM -- Department of Excellence 2023--27'' and by the funds that remained at the Vrije Universiteit Brussel at the completion of the ``QUARTZ (Quantum Information Access and Retrieval Theory)'' project, part of the ``Marie Sklodowska-Curie Innovative Training Network 721321'' of the ``European Unions Horizon 2020'' research and innovation program, with Diederik Aerts as principle investigator for the Brussels part of the network.

\end{document}